\documentclass[12pt,thmsa]{article}
\usepackage{sw20lart}

\input tcilatex
\begin{document}

\author{Amjad Hussain Shah Gilani \\
National Center for Physics\\
Quaid-i-Azam University\\
Islamabad 45320, Pakistan\\
Email: ahgilani@yahoo.com}
\title{Decay mechanism of quarks and leptons}
\date{}
\maketitle

\begin{abstract}
The top-family quarks will decay in charm-family quarks while beauty-family
quarks serve the purpose of spectrators. Decay matrix is suggested for quark
and lepton decays.
\end{abstract}

The Large Hadron Collider (LHC) is the global collaboration in particle
physics. The LHC will probe deeper into matter than ever before to explore a
new energy region and search for new phenomena. LHC is scheduled to start in
April 2007. The 27 kilo-meter rings of LHC will circulate two
counter-rotating beams of protons at nearly the speed of light. Before the
run of LHC in 2007, we have to explore the maximum possibilities on
theoretical grounds. Quantum Choromodynamics (QCD), the theory of strong
interactions, is unique among physical theories in its combination of
logical closure and empirical success. In the asymptotic Bjorken limit, the
study of the properties of strong interactions has proved to be one of the
most creative ideas \cite{hep-ph/0502113}. Gluons are responsible for the
strong interactions between the colored quarks and also interact themselves.
Gluons carry color-anticolor charge. In a recent study, a group property
violation in the charge structure of gluons \cite{hep-ph/0404026} is
observed. This violation in the charge structue of gluons changes the entire
scnario of the QCD but favours the electroweak theory \cite
{hep-ph/0404026,hep-ph/0410207,hep-ph/0501103,hep-ph/0502055,hep-ph/0502117,hep-ph/0503025,hep-ph/0503196}%
.

In the present study, we suspect that only top-family quarks will decay into
charm-family quarks while the beauty-family quarks will serve the purpose of
spectrator. We expect the following decay matrix

\[
V_{quarks}=\left( 
\begin{tabular}{cccccccc}
$V_{00}$ & $V_{0r}$ & $V_{0g}$ & $V_{0b}$ & $V_{0z}$ & $V_{0\bar{b}}$ & $V_{0%
\bar{g}}$ & $V_{0\bar{r}}$ \\ 
$V_{r0}$ & $V_{rr}$ & $V_{rg}$ & $V_{rb}$ & $V_{rz}$ & $V_{r\bar{b}}$ & $V_{r%
\bar{g}}$ & $V_{r\bar{r}}$ \\ 
$V_{g0}$ & $V_{gr}$ & $V_{gg}$ & $V_{gb}$ & $V_{gz}$ & $V_{g\bar{b}}$ & $V_{g%
\bar{g}}$ & $V_{g\bar{r}}$ \\ 
$V_{b0}$ & $V_{br}$ & $V_{bg}$ & $V_{bb}$ & $V_{bz}$ & $V_{b\bar{b}}$ & $V_{b%
\bar{g}}$ & $V_{b\bar{r}}$ \\ 
$V_{z0}$ & $V_{zr}$ & $V_{zg}$ & $V_{zb}$ & $V_{zz}$ & $V_{z\bar{b}}$ & $V_{z%
\bar{g}}$ & $V_{z\bar{r}}$ \\ 
$V_{\bar{b}0}$ & $V_{\bar{b}r}$ & $V_{\bar{b}g}$ & $V_{\bar{b}b}$ & $V_{\bar{%
b}z}$ & $V_{\bar{b}\bar{b}}$ & $V_{\bar{b}\bar{g}}$ & $V_{\bar{b}\bar{r}}$
\\ 
$V_{\bar{g}0}$ & $V_{\bar{g}r}$ & $V_{\bar{g}g}$ & $V_{\bar{g}b}$ & $V_{\bar{%
g}z}$ & $V_{\bar{g}\bar{b}}$ & $V_{\bar{g}\bar{g}}$ & $V_{\bar{g}\bar{r}}$
\\ 
$V_{\bar{r}0}$ & $V_{\bar{r}r}$ & $V_{\bar{r}g}$ & $V_{\bar{r}b}$ & $V_{\bar{%
r}z}$ & $V_{\bar{r}\bar{b}}$ & $V_{\bar{r}\bar{g}}$ & $V_{\bar{r}\bar{r}}$%
\end{tabular}
\right) 
\]
where $V_{00}=V_{c_0t_0}$, $V_{0r}=V_{c_0t_r}$, etc. So far we have
considered the most general posibilty of the decay mechanism of quarks. The
matrix $V$ is obtained on the basis of Cabbibo-Kobayashi-Maskawa (CKM)
matrix \cite{PRL10-531,PTP49-652}. We can suspect the similar decay
mechanism for the leptons and expect the decay matrix 
\[
U_{leptons}=\left( 
\begin{tabular}{cccccccc}
$U_{00}$ & $U_{0r}$ & $U_{0g}$ & $U_{0b}$ & $U_{0z}$ & $U_{0\bar{b}}$ & $U_{0%
\bar{g}}$ & $U_{0\bar{r}}$ \\ 
$U_{r0}$ & $U_{rr}$ & $U_{rg}$ & $U_{rb}$ & $U_{rz}$ & $U_{r\bar{b}}$ & $U_{r%
\bar{g}}$ & $U_{r\bar{r}}$ \\ 
$U_{g0}$ & $U_{gr}$ & $U_{gg}$ & $U_{gb}$ & $U_{gz}$ & $U_{g\bar{b}}$ & $U_{g%
\bar{g}}$ & $U_{g\bar{r}}$ \\ 
$U_{b0}$ & $U_{br}$ & $U_{bg}$ & $U_{bb}$ & $U_{bz}$ & $U_{b\bar{b}}$ & $U_{b%
\bar{g}}$ & $U_{b\bar{r}}$ \\ 
$U_{z0}$ & $U_{zr}$ & $U_{zg}$ & $U_{zb}$ & $U_{zz}$ & $U_{z\bar{b}}$ & $U_{z%
\bar{g}}$ & $U_{z\bar{r}}$ \\ 
$U_{\bar{b}0}$ & $U_{\bar{b}r}$ & $U_{\bar{b}g}$ & $U_{\bar{b}b}$ & $U_{\bar{%
b}z}$ & $U_{\bar{b}\bar{b}}$ & $U_{\bar{b}\bar{g}}$ & $U_{\bar{b}\bar{r}}$
\\ 
$U_{\bar{g}0}$ & $U_{\bar{g}r}$ & $U_{\bar{g}g}$ & $U_{\bar{g}b}$ & $U_{\bar{%
g}z}$ & $U_{\bar{g}\bar{b}}$ & $U_{\bar{g}\bar{g}}$ & $U_{\bar{g}\bar{r}}$
\\ 
$U_{\bar{r}0}$ & $U_{\bar{r}r}$ & $U_{\bar{r}g}$ & $U_{\bar{r}b}$ & $U_{\bar{%
r}z}$ & $U_{\bar{r}\bar{b}}$ & $U_{\bar{r}\bar{g}}$ & $U_{\bar{r}\bar{r}}$%
\end{tabular}
\right) ,
\]
where $V_{00}=V_{e_0\tau _0}$, $V_{0r}=V_{e_0\tau _r}$, etc. For the
predictions, we will follow the calculational details given in Refs. \cite
{IJMPA17-4927,hep-ph/0107330}.

\end{document}